\documentclass[floatfix,twocolumn,aps,prd,showpacs,amsmath,amssymb]{revtex4-1}
\usepackage[latin1]{inputenc}
\usepackage[dvips]{graphicx}

\usepackage{dcolumn}
\usepackage{bm}
\usepackage{dsfont}
\usepackage{color}
\usepackage{ulem} 
\usepackage{url}
\usepackage[colorlinks=true ,citecolor=blue]{hyperref}
\usepackage{amsmath,amssymb}

\begin{document}

\title{Momentum-Space Cigar Geometry in Topological Phases}

\author{Giandomenico Palumbo}
\affiliation{$^1$Institute for Theoretical Physics, Center for Extreme Matter and Emergent Phenomena, Utrecht University, Princetonplein 5, 3584CC Utrecht, the Netherlands}

\date{\today}

\begin{abstract}
In this paper, we stress the importance of momentum-space geometry in the understanding of two-dimensional topological phases of matter. We focus, for simplicity, on the gapped boundary of three-dimensional topological insulators in class AII, which are described by a massive Dirac Hamiltonian and characterized by an half-integer Chern number. The gap is induced by introducing a magnetic perturbation, such as an external Zeeman field or a ferromagnet on the surface. The quantum Bures metric acquires a central role in our discussion and identifies a cigar geometry. We first derive the Chern number from the cigar geometry and we then show that the quantum metric can be seen as a solution of two-dimensional non-Abelian BF theory in momentum space. The gauge connection for this model is associated to the Maxwell algebra, which takes into account the Lorentz symmetries related to the Dirac theory and the momentum-space magnetic translations connected to the magnetic perturbation. The Witten black-hole metric is a solution of this gauge theory and coincides with the Bures metric. This allows us to calculate the corresponding momentum-space entanglement entropy that surprisingly carries information about the real-space conformal field theory describing the defect lines that can be created on the gapped boundary.
 \end{abstract}

\maketitle

\section{Introduction}
Geometry is nowadays recognized as a fundamental ingredient in understanding of topological phases of matter \cite{1}. Several physical properties, such as Hall viscosity \cite{2,3} and thermal Hall effect \cite{Kane,4,5}, can be naturally de- scribed through effective geometric theories. Most of the efforts have been directed to the formulation of geometric models in real space. Deformations and dislocations in the lattice \cite{6,7} and interactions in continuous systems \cite{8} can be represented by effective models in real curved space. Some of these models are characterized, for instance, by a non-minimal coupling between gauge field and background \cite{9,10}, non-zero torsion \cite{6,11}, quantum anomalies \cite{12,13,14}, spin-orbit coupling in curved space \cite{15}, etc.

In an our recent work \cite{10}, we have employed a geometric Chern-Simons theory in real space based on a 2+1-dimensional Maxwell algebra \cite{16,17} to study the gapped boundary of three-dimensional topological insulators in class AII \cite{18}. This algebra encodes both Lorentz symmetries and magnetic translations \cite{19,20}. The former are related to the Dirac surface states while the latter are due to the presence of a magnetic perturbation, such as an external Zeeman field or ferromagnet \cite{21}, that induces a gap on the surface.

However, the role of momentum-space geometry is much less understood in topological materials. A real-space geometric theory cannot be naturally translated in a momentum-space model due to the lacking of a well-defined Fourier transformation in a generic curved background. Besides the well-known Berry phase, the quantum Bures (or Fubini-Study) metric can be derived from the band structures of lattice models \cite{22,23}. It represents the natural metric of an effective curved momentum-space in condensed matter systems. Recent works point out that this kind of geometry has deep physical implications in topological phases \cite{24,25,26,27,28,29,30,31,32}. Similar ideas about curved momentum-space geometry have been also developed in high-energy physics literature \cite{33,34,35,36,37,38,39}.

\begin{figure}[!htb]
	\centering
	\includegraphics[width=0.42\textwidth]{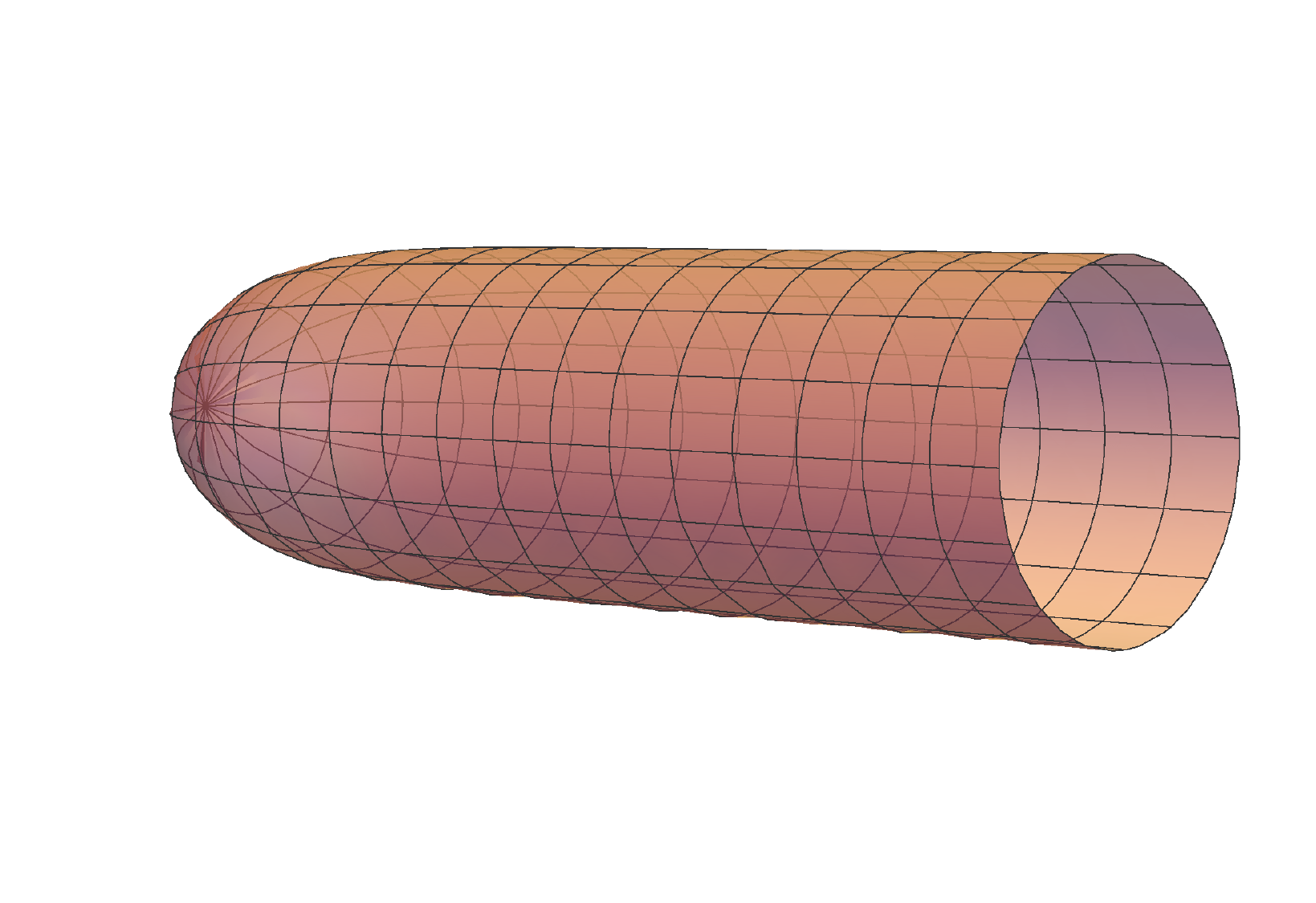}
	\caption{The momentum-space geometry of a massive Dirac Hamiltonian is given by a cigar geometry. For small momenta, the geometry approaches a semi-sphere, while for large momenta becomes a cylinder. It is characterized by a Bures metric that coincides with the Witten black-hole metric. This black hole has the horizon at the tip of the cigar.}
	\label{System}
\end{figure}

In this paper, we focus on the geometric properties of the momentum space of two-dimensional topological phases. In particular, we analyze the gapped boundary of three-dimensional topological insulators in class AII, which is described by a massive Dirac Hamiltonian \cite{18}. In this case, the Bures metric identifies a cigar geometry as shown by Matsuura and Ryu \cite{32}. We firstly show that the a Chern number can be directly derived from the cigar geometry and coincides with the value calculated through the standard Berry phase. We then propose a suitable non-Abelian BF theory \cite{40} in the two-dimensional momentum space, where the Witten black-hole (WBH) metric \cite{41,42} is a solution of its equations of motion and coincides with the Bures metric of the cigar geometry. The gauge connection of the BF theory is related to the lower-dimensional Maxwell algebra, employed in the study of dilaton gravity \cite{40} and conformal field theories (CFTs) \cite{43}. This algebra is basically a central extension of the two-dimensional Euclidean Poincar\'e algebra. The choice of this gauge theory is physically well justified. Firstly, the BF theory represents a non-trivial topological field theory in two dimensions \cite{44}. Different versions of this theory have been successfully employed in the study of topological insulators, superconductors, superfluids and graphene in real flat space \cite{45,46,47,48,49,50}. Secondly, the Maxwell algebra, naturally encodes the Lorentz symmetries related to the Dirac theory and the momentum-space magnetic translations \cite{27} connected to the magnetic perturbation. Finally, by following the results by Solodukhin \cite{51,52} on the entanglement entropy (EE) of the WBH, we show that the momentum-space EE contains information about the CFT related to the topological phase. This critical theory describes the defect lines that be created on the gapped boundary of the three-dimensional topological insulator. Our results enlighten novel and important connections between geometric and topological features of condensed-matter systems.

\section{Bures metric and Chern number}

We start by stressing the role of gauge formalism in the momentum space of topological systems. Several topological invariants associated to the band structure are indeed related to Abelian and non-Abelian Berry phase, which depends on the Bloch wave-functions \cite{1}. For instance, in two-dimensional Chern insulators, the integral of the U(1) Berry curvature on the first Brillouin zone, gives us the Chern number \cite{53}. In three-dimensional time-reversal-invariant topological insulators, the corresponding Z$_{2}$ topological number is given in terms of the non-Abelian Chern-Simons invariant of a non-Abelian Berry connection [18]. All these topological terms are defined in terms of gauge connections, which are static gauge fields, i.e. they cannot be solutions of any effective action. Intriguously, it has been recently proposed a dynamical version of the Berry connection in three-dimensional topological superconductors, that is a solution of the equations of motion of a non-Abelian Chern-Simons theory \cite{54}. BF theory \cite{44} represents the natural topological field theory in two dimensions.

In the next section, we will show that a non-Abelian BF theory \cite{40} can describes the momentum-space geometry of a suitable two-dimensional topological phase, given in terms of the Bures metric $g_{\mu\nu}$. This metric is the real part of the quantum geometric tensor $G_{\mu\nu}(\textbf{k})$ \cite{22,23}
\begin{eqnarray}
G_{\mu\nu}(\textbf{k})=g_{\mu\nu}(\textbf{k})+\dfrac{i}{2} F_{\mu\nu}(\textbf{k}),
\end{eqnarray}
where the Abelian Berry curvature $F_{\mu\nu}$ represents its imaginary part. The Bures metric is explicitly given by
\begin{eqnarray}
g_{\mu\nu}(\textbf{k})=\dfrac{1}{8} {\rm tr}[\partial_{\mu}Q\partial_{\nu}Q],
\end{eqnarray}
with $\mu,\nu=k_{x},k_{y},...,k_{j}$ and
\begin{eqnarray}
Q(\textbf{k})=1-2 P(\textbf{k}), \hspace{0.4cm} P(\textbf{k})=\sum_{i}|u_{i}(\textbf{k})\rangle \langle u_{i}(\textbf{k})|,
\end{eqnarray}
where $P(\textbf{k})$ is the spectral projector onto the filled Bloch wave-functions $u_{i}(\textbf{k})$. The quantum metric plays an important role in many-body systems and generally carries different information with respect to the Berry phase. The Bures metric has been connected to physical properties and observables of two-dimensional systems, such as density operators \cite{24,25,26}, quantum phase transitions \cite{55,56,57}, superfluid weight \cite{58}, orbital susceptibility \cite{59,60}.
Here, we focus on the geometric properties of gapped boundary of three-dimensional insulators in class AII, where a gap is induced by an external Zeeman field or by an ferromagnet on the surface \cite{21}. For a single Dirac mode, the surface state is described in the low-energy regime by a two-dimensional massive Dirac Hamiltonian
\begin{eqnarray}\label{Dirac}
H= \sigma_{x}k_{x}+\sigma_{y}k_{y}+m \sigma_{z},
\end{eqnarray}
where $\sigma_{j}$ are the Pauli matrices, $m$ is the Dirac mass induced by the magnetic perturbation and the Fermi velocity is fixed to $1$. By assuming $m > 0$, the corresponding Berry connection and curvature for the lowest filled band $u(\textbf{k})$ are respectively given by
\begin{eqnarray}
A_{x}(\textbf{k})=\langle u|\partial_{k_{x}} |u \rangle=\dfrac{i k_{y}}{2\lambda(\lambda+m)}, \nonumber \\
A_{y}(\textbf{k})=\langle u|\partial_{k_{y}} |u \rangle=\dfrac{i k_{y}}{2\lambda(\lambda+m)},
\end{eqnarray}
\begin{eqnarray}
F_{xy}(\textbf{k})=\partial_{k_{x}}A_{y}-\partial_{k_{y}}A_{x}=-\frac{i m}{2 \lambda^{3}}.
\end{eqnarray}
The topological invariant at low momenta ($k \approx 0$) is the first Chern number $C_{1}$
\begin{eqnarray}\label{Chern}
C_{1}=\frac{i}{2\pi}\int_{BZ} d^{2}k F_{xy}=\frac{1}{2}\frac{m}{|m|},
\end{eqnarray}
which characterizes an half-integer quantum Hall state on each gapped surface. The quantum metric for the Hamiltonian in Eq. (\ref{Dirac}) has been derived in Ref. \cite{32}
\begin{eqnarray}
g_{\mu\nu}(\textbf{k})=\frac{1}{4}\left( \frac{m^{2}}{\lambda^{4}} dk^{2}+\frac{k^{2}}{\lambda^{2}} d\theta^{2}\right) 
\end{eqnarray}
where $k^{2} =k_{x}^{2}+k_{y}^{2}$, $\lambda=\sqrt{k^{2}+m^{2}}$ and $d\theta$ is the line element of the unit circle in polar coordinate. In particular, we have that
\begin{eqnarray}
g_{\mu\nu}(\textbf{k})|_{k\rightarrow 0}=\frac{1}{4 m^{2}}(dk^{2}+k^{2}d\theta^{2}), \nonumber \\
g_{\mu\nu}(\textbf{k})|_{k\rightarrow \infty}=\frac{1}{4}(m^{2}d\xi^{2}+d\theta^{2}), \hspace{0.3cm}
\end{eqnarray}
where $\xi=1/k$. This implies that the effective momentum space is a cigar geometry, which has positive Gaussian curvature at low momenta and is asymptotically equivalent to a cylinder with finite circumference at large momenta, see Fig.1.
We now employ K\"ahler geometry \cite{Nakahara} to show the deep relation between the topological information contained in the cigar geometry and the first Chern number associated to the Berry phase. We can indeed complexify our momentum space by introducing the following complex variables
\begin{eqnarray}
z=k_{x}+i k_{y}, \hspace{0.4cm} \bar{z}=k_{x}-i k_{y},
\end{eqnarray}
such that the cigar, up to some normalized coefficient, is defined by the following infinitesimal line element $ds^{2}$ \cite{Chow}
\begin{eqnarray}
ds^{2}=g_{\mu\bar{\nu}} dz^{\mu}d\bar{z}^{\nu}=\frac{dz d\bar{z}}{1+z \bar{z}},
\end{eqnarray}
where $g_{\mu\bar{\nu}}$ a K \"ahler metric. Notice, that complex spaces have been already applied in the study of Majorana fermions and topological phases \cite{Hatsugai,Mandal, Chua}. It is then possible to calculate the Ricci curvature $\mathfrak{R}_{\mu\bar{\nu}}$ and the Ricci form $\mathfrak{R}$ of the above metric, respectively given by
\begin{eqnarray}
\mathfrak{R}_{\mu\bar{\nu}}=-\partial_{\mu}\partial_{\bar{\nu}} \ln \det (g_{\mu\bar{\nu}}), \nonumber \\
\mathfrak{R}= \mathfrak{R}_{\mu\bar{\nu}} dz^{\mu}\wedge d\bar{z}^{\nu}, \hspace{0.65cm}
\end{eqnarray}
where $\det (g_{\mu\bar{\nu}})=-1/(1+z \bar{z})^{2}$ and $\wedge$ is the wedge product.
We can now employ an important theorem in K\"ahler geometry that states that the Chern curvature form ${\rm Ch}$ associated to the complex tangent bundle is proportional to the Ricci form \cite{Nakahara}, namely
\begin{eqnarray}
{\rm Ch}=\frac{\mathfrak{R}}{2\pi}.
\end{eqnarray}
The Chern connection coincides with the complexification of Levi-Civita connection of the (Riemannian) metric on the underlying real manifold (i.e. the real part of the K\"ahler metric).
The integral of the Chern curvature form on a compact complex manifold gives us always an integer topological invariant $C_{1}$, which is nothing but the first Chern number. However, because the cigar is non-compact, the integral of the Ricci form gives us $C_{1} = 1/2$ due to the positive curvature of the cigar cap (which is a semi-sphere) at small momenta $k$. This is perfectly in agreement with the derivation of the half-integer Chern number from the integral of the Berry curvature in Eq. (\ref{Chern}). This equivalence between the Ricci form and the Berry curvature form has been previously remarked in Ref. \cite{Page}.

\section{BF theory from the Maxwell algebra}

In this section, we will show that the quantum metric that identifies the cigar geometry can be seen as a solution of the equations of motion of a BF theory in momentum space. Even if the Bures metric is completely fixed by the band structure, nevertheless the existence of an effective action is rlevant because it contains always more information than its equations of motion. We first have to fix the underlying gauge algebra of momentum space. Like in the real-space geometry \cite{10}, the algebra has to take into account the Lorentz symmetries related to the Dirac fermions and the momentum-space magnetic translations \cite{27} connected to the Berry connection, which behaves like and electromagnetic potential. This is the signature of position-momentum duality in quantum Hall states as recently emphasized in Refs. \cite{27,Carusotto}. For these reasons, we consider the Maxwell algebra, which is identified by the following anti-commuting relations
\begin{eqnarray}
[P_{a},P_{b}]=\epsilon_{ab} Z, \hspace{0.3cm} [P_{a},J]=\epsilon_{a}^{\,\,b}P_{b}, \hspace{0.3cm} [Z,P_{a}]=0,
\end{eqnarray}
where $a, b = 1, 2$, $P_{a}$ are the generators of the translations, $J$ is the generator of the Lorentz rotations and $Z$ is related to the central extension of the algebra. The corresponding gauge connection $\mathcal {A}_{\mu}$ is given by
\begin{eqnarray}
\mathcal {A}_{\mu}=\frac{1}{\beta} e_{\mu}^{a} P_{a}+\omega_{\mu}J+a_{\mu} I,
\end{eqnarray}
where $\mu,\nu = k_{x}, k_{y}$, $\beta$ is a dimensionful constant, $\omega_{\mu}$ represents a momentum-space spin connection, $e_{\mu}^{a}$ is a momentum-space zweibein, such that $g_{\mu\nu} = e_{\mu}^{a}e_{\nu}^{b}\delta_{ab}$, and $a_{\mu}$ is an effective U(1) potential. The curvature tensor acquires the following form
\begin{eqnarray}
\mathcal {F}_{\mu\nu}=\frac{1}{\beta}(D_{\mu}e_{\nu}^{a}-D_{\nu}e_{\mu}^{a})P_{a}+(\partial_{\mu}\omega_{\nu}-\partial_{\nu}\omega_{\mu})J+
\nonumber \\ \left(\partial_{\mu}a_{\nu}-\partial_{\nu}a_{\mu}+\frac{1}{2 \beta^{2}} \epsilon_{ab}\, e^{a}_{\mu}e^{b}_{\nu} \right) I, \hspace{0.9cm}
\end{eqnarray}
where $D_{\mu} = \partial_{\mu} + \omega_{\mu}$ is the (Abelian) covariant derivative. Because we are interested in the topological and geometric responses of a topological phase, we consider the action of a topological field theory, given by the following non-Abelian BF action
\begin{eqnarray}
\mathcal{S}_{{\rm BF}}[B,\mathcal{A}_{\mu}]=\int d^{2}k\, \epsilon^{\mu\nu} B_{\hat{a}}\mathcal{F}_{\mu\nu}^{\hat{a}},
\end{eqnarray}
where $\hat{a}=1,2,3,4$, $B_{\hat{a}}$ are some scalar fields (Lagrange multipliers) and ${F}_{\mu\nu}^{\hat{a}}$ are the components of the curvature tensor. It can be explicitly written in terms of spin connection, zweibein and effective gauge potential as follows
\begin{align}\label{BF}
\mathcal{S}_{{\rm BF}}=\int \left[  B_{a}(D e)^{a}+B_{3}d\omega+B_{4}\left(da+\frac{1}{2} \epsilon_{ab}\,e^{a}e^{b} \right) \right],
\end{align}
where the constant $\beta = 1$, which plays the role for momentum-space magnetic length \cite{Carusotto}, has been fixed for simplicity. By replacing the zweibein with the metric tensor, it is then possible to rewrite the above action in the second-order formalism (up to a total derivative) \cite{40}
\begin{eqnarray}
\mathcal{S}[\phi,g_{\mu\nu}]=\int d^{2}k\, \sqrt{g}\,(\phi R-\Lambda),
\end{eqnarray}
where $g=\det (g_{\mu\nu})$, $R=(1/\sqrt{g})d\omega$ is the Ricci scalar, $\phi=B_{3}$ and $\Lambda=-2B_{4}$ is a constant.
This action is conformally equivalent to that one proposed in Refs. \cite{41,42,Callan} in the context of dilaton gravity
\begin{eqnarray}\label{dilaton}
\mathcal{S}[\varphi,\bar{g}_{\mu\nu}]=\int d^{2}k\, \sqrt{\bar{g}}\,(\bar{R}-4 \bar{g}^{\mu\nu}\partial_{\mu}\varphi\partial_{\nu}\varphi-\Lambda),
\end{eqnarray}
where $\phi=e^{-2\varphi}$ and $\bar{g}_{\mu\nu}=e^{2\varphi}g_{\mu\nu}$. As shown in Refs. \cite{41,42}, the (Euclidean) WBH metric is a solution of the equations of motion of Eq. (\ref{dilaton}) and characterizes a cigar geometry, see Fig.1. In terms of the infinitesimal line element $ds^{2}$, this metric is given by
\begin{eqnarray}
ds^{2}=\frac{1}{f(k_{x})}dk_{x}^{2}+f(k_{x})dk_{y}^{2},
\end{eqnarray}
where $f(k_{x})=1-e^{-\Omega k_{x}}$, with $\Omega$ a constant related to the thermodynamics of the effective black hole \cite{51,52}.
Here, $k_{y}$ plays the same role of Euclidean time in real-space Euclidean black holes. This kind of metric has been intensively studied in high-energy physics and has a natural relation with the area-preserving diffeomorphisms and $W_{\infty}$ algebra \cite{Mavromatos}. These mathematical structures are also essential ingredients in the quantum Hall effect \cite{20}.

As a final point in this section, we calculate the stress-energy tensor $T_{\mu}^{a}$ of the model by varying the action in Eq. (\ref{BF}) with respect to $e_{\mu}^{a}$
\begin{eqnarray}\label{stress-energy}
T_{\mu}^{a}=D_{\mu}B^{a}+B_{4}\epsilon_{b}^{\,\,a}e^{b}_{\mu}.
\end{eqnarray}
We can compare this result with the elastic response of the system in momentum space. Firstly, like in real space, the interpretation of the zweibein in terms of elasticity quantities, i.e. the displacement field $\chi^{a}$ \cite{6}, tells us that around the flat-space configuration, we have that
\begin{eqnarray}
e_{\mu}^{a}=\delta_{\mu}^{a}+w_{\mu}^{a},
\end{eqnarray}
where $w_{\mu}^{a}=\partial_{\mu}\chi^{a}$ is the distorsion tensor. Secondly, $w_{\mu}^{a}$ carries information about the line defects (dislocations) through the calculation of the line-integral
\begin{eqnarray}
\oint w_{\mu}^{a} dk^{\mu} =-b^{a},
\end{eqnarray}
where $b^{a}$ is the Burgers vector. Thus, the relation in Eq. (\ref{stress-energy}) implies that the stress-energy tensor contains information about possible nodal lines in momentum space. In gapped topological phases, these kinds of defects appear, for instance, in a recently proposed Chern insulator with a nonsymmorphic symmetry \cite{Zhou}.

\begin{figure}[!htb]
	\centering
	\includegraphics[width=0.40\textwidth]{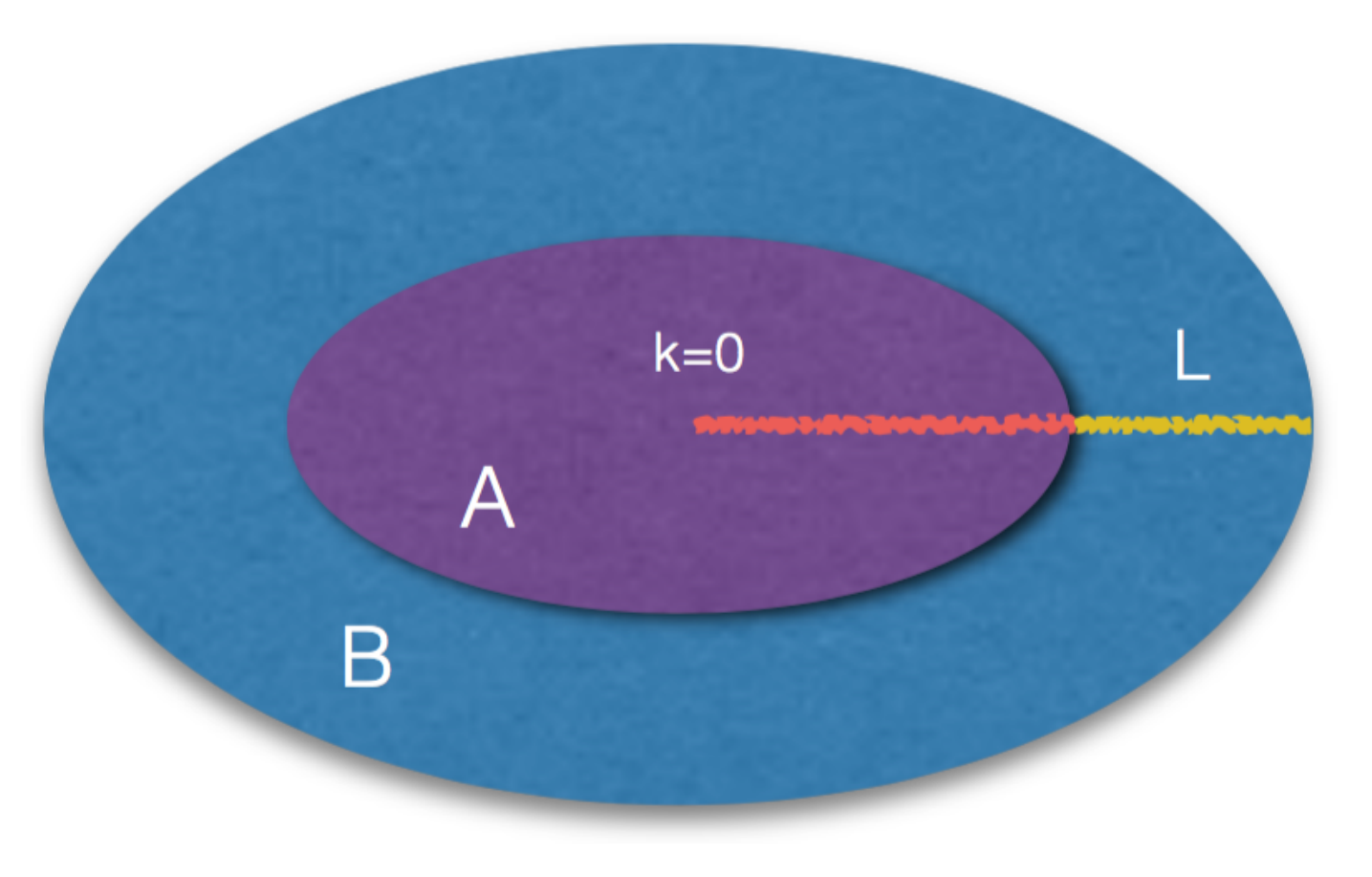}
	\caption{The tip of the cigar at $k = 0$ coincides with the center of the disk, being the cigar and the disk conformally equivalent. The momentum space is divided in two sub-regions A and B, with the former containing the lower-energy modes. A radial line is represented by a red-yellow line. The length L defines an interval between the boundaries of A and B.}
	\label{System}
\end{figure}

\section{Momentum-space entanglement entropy}

We will describe here the main implications of cigar geometry in the context of momentum-space entanglement entropy (EE). In particular, we will show that a bipartition of a radial line in momentum space contains information about the CFT related to the topological phase on the gapped boundary.
CFTs and topological phases have been intensively analyzed by employing the tools of quantum information, such as EE \cite{1} and entanglement spectrum \cite{Haldane2}. To define the former, we first bipartite the Hilbert space of a given quantum system with density matrix $\rho$, into two non-overlapping regions A and B. We then quantify the entanglement between the two regions by tracing out, for instance, the degrees of freedom in B. In such a way, the corresponding EE is given by
\begin{eqnarray}
S_{{\rm EE}}=-{\rm tr}\, \rho_{A} \ln \rho_{A},
\end{eqnarray}
where $\rho_{A}={\rm tr}_{B}\rho$ is the reduced density matrix. Recently, these approaches have been generalized to momentum space \cite{Bala,Hughes,Lungren,Qi2}. In this case, the separation of the degrees of freedom is related to the energy scale in analogy to the Wilson's renormalization group. Here, the infrared (IR) regime of a quantum system is studied by integrating out the ultraviolet (UV) degrees of freedom. As shown in Ref. \cite{Bala}, the momentum-space entanglement entropy for IR degrees of freedom is defined by tracing over the UV degrees of freedom with momenta $k>\mu$, where $\mu$ is a fixed energy scale.
In topological systems, a real-space entanglement cut carries information about the CFT of the edge states \cite{Bernevig,Haldane2}. By following this idea, we show how a radial line together with the momentum-space EE provide us information about the CFT related to the topological phase.

We remind that the WBH has a well defined Hawking temperature $T_{{\rm H}}$ given by \cite{51}
\begin{eqnarray}
T_{{\rm H}}=\frac{\Omega}{4\pi}.
\end{eqnarray}
Moreover, the cigar is conformally equivalent to a flat disk, see Fig.2. We then divide the system in two regions A and B, where the lower momenta ($k \approx 0$) are contained in the region A. Due to the rotational symmetry of the disk, we introduce a radial line, which is bipartited in two linear sub-regions. We call $L$ the length of the line that lies in the sub-region B. It is geometrically expressed as follows 
\cite{51}
\begin{eqnarray}\label{len}
L=\int_{L_{A}}^{L_{B}} dk_{x}\,\frac{1}{\sqrt{1-e^{-\Omega k_{x}}}},
\end{eqnarray}
where $L_{A}$ and $L_{B}$ are the boundary of regions A and B, respectively. In black holes, there exists a natural bipartition of the space due to the presence of an horizon. However, in the WBH, the horizon is at the tip of the cigar, i.e. at $k \approx 0$ because $f(k_{x})$ has a zero at $k_{x} = 0$. This implies that the integral in Eq.(\ref{len}) has to be considered in the limit $L_{A} \rightarrow 0$. We remind also that thisi horizon is zero-dimensional, being the boundary of the one-dimensional space-like region, which is the radial line in our case.
The EE of the cigar geometry has been already calculated by Solodukhin \cite{51,52}
\begin{eqnarray}\label{entropy}
S_{{\rm cigar}}^{EE}=\dfrac{c}{6} \ln \left( \dfrac{1}{2\pi T_{{\rm H}}\varepsilon}\sinh(2\pi T_{{\rm H}} L)\right),
\end{eqnarray}
where $\varepsilon$ is a UV regulator and $c = 1$ is the central charge. This expression coincides with the finite-temperature EE of a free CFT on a semi-infinite line \cite{Cardy}.
Moreover, for  $T_{{\rm H}}\rightarrow 0$, we recover the corresponding zero-temperature EE
\begin{eqnarray}
S_{{\rm cigar}}(T_{{\rm H}}\rightarrow 0)=\dfrac{c}{6} \ln \left(\frac{L}{\varepsilon}\right).
\end{eqnarray}

In the thermodynamic limit, i.e. for $L$ very large, from Eq. (\ref{entropy}), we obtain the thermal entropy $S_{{\rm cigar}}^{th}$
\begin{eqnarray}
S_{{\rm cigar}}^{th}=\dfrac{\pi c}{3}L\, T_{{\rm H}},
\end{eqnarray}
which coincides with the thermal entropy of a one-dimensional massless Dirac fermion when $T=v T_{{\rm H}}$ \cite{Affleck, Cardy2}, where $v$ is the Fermi velocity of the Dirac mode.

Thus, we have shown that the dual of the real-space entanglement entropy of the CFT emerges from the momentum-space geometry of a topological phase. This CFT is nothing but the critical theory associated to the defect lines that can be created on the topological gapped boundary. Along each defect, a single chiral Dirac mode propagates. This result is also in agreement with the position-momentum duality of the EE as already emphasized in Ref. \cite{Qi2}.

\section{Conclusions}

By summarizing, we have analyzed the momentum-space geometry of the gapped boundary of three-dimensional topological insulators. By starting from the corresponding momentum-space cigar geometry, we have derived the Chern number in a purely geometric way. We have then shown that the momentum-space metric, formally equivalent to the metric of a lower-dimensional black hole, can be seen as a semi-classical solution of a non-Abelian BF theory. Finally, we have calculated the momentum-space EE following the approach employed in black-hole physics. This entropy is equivalent to the EE of a CFT at finite temperature in flat space. It carries information about the gapless chiral Dirac modes that appear along the defect lines on the gapped topological boundary. This represents a further evidence of the position-momentum duality of the entanglement in topological systems. Our results strengthen the importance of momentum-space geometry through gauge theory in topological phases, providing novel research directions in the study of more complex quantum many-body systems.

\acknowledgments
We thank Andrea Cappelli and Miguel Montero for discussions and comments.

\end{document}